# Quasar Rain: the Broad Emission Line Region as Condensations in the Warm Accretion Disk Wind

Martin Elvis

Harvard-Smithsonian Center for Astrophysics

## ABSTRACT

The origin of the broad emission line region (BELR) in quasars and active galactic nuclei is still unclear. I propose that condensations form in the warm, radiation pressure driven, accretion disk wind of quasars creating the BEL clouds and uniting them with the other two manifestations of cool (~$10^4$ K) gas in quasars, the low ionization phase of the warm absorbers (WAs) and the clouds causing X-ray eclipses. The cool clouds will condense quickly (days – years), before the WA outflows reach escape velocity (which takes months – centuries). Cool clouds form in equilibrium with the warm phase of the wind because the rapidly varying X-ray quasar continuum changes the force multiplier, causing pressure waves to move gas into stable locations in pressure-temperature space. The narrow range of 2-phase equilibrium densities may explain the (luminosity)$^{1/2}$ scaling of the BELR size, while the scaling of cloud formation timescales could produce the Baldwin effect. These dense clouds have force multipliers of order unity and so cannot be accelerated to escape velocity. They fall back on a dynamical timescale (months - centuries), producing an inflow that rains down toward the central black hole. As they soon move at Mach ~10 – 100 with respect to the WA outflow, these "raindrops" will be rapidly destroyed within months. This rain of clouds may produce the elliptical BELR orbits implied by velocity resolved reverberation mapping in some objects, and can explain the opening angle and destruction timescale of the narrow "cometary" tails of the clouds seen in X-ray eclipse observations. Some consequences and challenges of this "quasar rain" model are presented along with several avenues for theoretical investigation.

## 1. INTRODUCTION

The broad (FWHM~5000 km s$^{-1}$, ~$0.02c$) emission lines (BELs) that dominate the optical and ultraviolet (UV) spectra of quasars and active galactic nuclei (AGNs; hereinafter both simply called "quasars" for convenience) are a puzzle. Although BELs were the first striking feature of quasars to be found, after their radio emission (Schmidt 1963), they still have no widely accepted explanation.

BELs appear to be a very common signature of quasars and so should be created by some robust physics. They are, by definition, found in all type 1 (broad-lined) quasars, e.g. the Sloan Digital Sky Survey (SDSS) quasar sample (Pâris et al. 2014). Although in the type 2 (narrow-line) objects the BELs are not evident they often appear in scattered-light polarized spectra (Antonucci & Miller 1985, Nagao et al., 2004 and references therein), and so are hidden from our view by dust obscuration. Similarly, in blazars a relativistically beamed jet pointed toward us can overwhelm the unbeamed BEL signal, but BELs are sometimes revealed in low state blazars (Pian et al. 1999, Giommi et al., 2012, Isler et al. 2015). So the BEL emitting region (BELR[1]) is likely a universal signature of supermassive black holes (M >~ $10^{5.5}$ $M_\odot$, Chakravorty et al. 2012) that are accreting at a sizeable fraction of the Eddington limit (>1%), as in quasars (Kollmeier et al., 2006, Steinhardt & Elvis, 2010).

Yet, after more than 50 years of research, the physical origin of the broad emission line region (BELR) is unclear. Although some accretion disk wind models (e.g. Murray et al., 1995, hereinafter MCGV) can make BELs from continuous media, most models involve large numbers of small clouds. The smoothness of BEL profiles supports continuous models (Laor et al. 2006), while X-ray eclipses by material with BELR properties (Risaliti et al., 2009, 2011, and see Section 2.2) strongly suggest that well-defined discrete clouds are present in the BELR. If so, then the number of clouds must be large, >$10^8$ (Arav et al. 1998). There have been numerous suggestions to explain these clouds.

One class of models involves the host galaxy, either employing stars, stellar clusters or supernovae (Alexander & Netzer 1994, Scoville & Norman 1988, Vilkovskij and Czerny 2002, Pittard et al., 2001), or utilizing accretion from large scales (Fromerth & Melia 2001, Gaspari Brighenti & Temi 2015). However, BEL properties (i.e. their ionic species, line widths and strengths) are quite universal. See, e.g., the similarity of the z=7.1 quasar to the mean SDSS spectrum (Mortlock et al. 2011, see also Baskin, Laor and Stern, 2014ab). This uniformity argues against any model that relies on large-scale galaxy properties beyond the black hole sphere of influence (Kormendy and Ho, 2013) at ~$10^5$ – $10^6$ Schwarzschild radii, $r_g$.

The other class of explanation for the BELR involves smaller scale properties more associated with the black hole: magnetic confinement (Rees 1987, Emmering, Blandford & Schlosman 1992), quasar winds (Blumenthal & Mathews 1979, Perry & Dyson 1985, Murray & Chiang 1998), radiation pressure confinement (Baskin, Laor & Stern, 2013, 2014abc), and thermal instabilities (Krolik, McKee & Tarter 1981, Krolik 1988, Mathews & Doane 1990, Proga & Waters 2015, Waters and Proga 2016).

Here I suggest a physical origin for the BELR that combines winds and thermal instabilities. The proposal is sketched in Figure 1. Briefly:

(1) Cool (~$10^4$ K) condensations with $10^{22} < N_H < 10^{26}$ cm$^{-2}$ will form quickly in an ouflowing warm (~$10^6$ K), radiatively driven, wind from the accretion disk, before the wind reaches escape velocity.
(2) Rapid changes in the X-ray/UV ratio of the ionizing continuum will drive these condensations to the quite restricted thermally stable, multi-phase, locations in pressure-temperatures space; the clouds will all then have effectively constant density. The clouds have the right physical conditions to be BELR clouds.
(3) The cool condensed clouds have ~100 times higher density than the warm medium from which they formed, and so an order of magnitude lower force multiplier for line driving of order unity. As a result they stall and fall back, creating an inflow. The condensations do not fall from infinity and will have only the angular momentum of the disk radius from which the wind gas is launched.

---
[1] As there are also broad absorption lines (BALs, e.g. Hall et al., 2002), this acronym is less ambiguous than the usual "BLR".

(4) The clouds are quite rapidly destroyed by ablation due to their highly supersonic (Mach number, $10 < M < 100$) motion through the warm wind, and so do not fall far or acquire high infall velocities.

(5) Their large transverse motions produce asymmetrical X-ray eclipses due to the small opening angles of their Mach cones.

These "raindrops" form the BELR clouds, the X-ray eclipsers and the low ionization phase (LIP) of the X-ray/UV warm absorbers (WAs). This proposal thus unifies the three cool (~$10^4$K) gas components in AGN.

This model will work only if the condensations form in the wind before reaching escape velocity, and if they are destroyed before falling far or impacting the accretion disk. Section 2 presents, first theoretical reasons to expect cool clouds to form, and then observational evidence for the presence of cool clouds having common physical conditions in each of the three types of cool cloud in AGN. Section 3 discusses the model in more detail, presenting the relevant timescales for cloud formation and destruction. Several consequences of this model are presented in Section 4. Some challenges to the model are given in Section 5.

Rain occurs when condensations in a gas can no longer be supported against gravity. "Quasar rain" then seems like an appropriate name for this infalling BEL material.

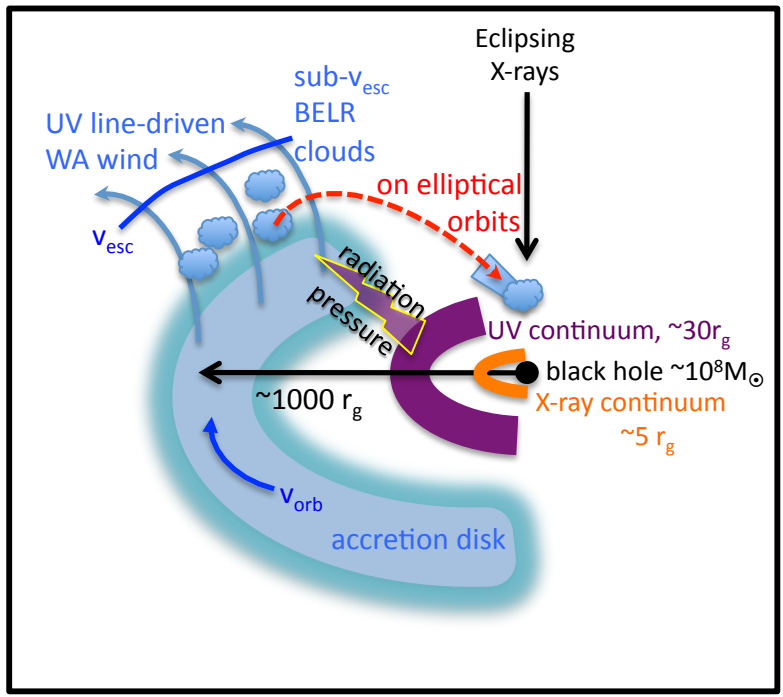

Figure 1. Sketch of the Quasar Rain scenario

## 2. COOL PHASES IN WARM AGN WINDS

If instabilities in an accretion disk wind are to create BEL clouds then a multi-phase wind is needed. It happens that it is both theoretically and observationally reasonable to expect a multi-phase medium in the X-ray and UV absorbing "warm absorber" (WA) winds of AGN for which the physical conditions in the coolest (~$10^4$ K), low ionization, phase are those of BELR gas.

### *2.1 Theoretical*

Krolik, McKee & Tarter (1981) showed that gas irradiated by a quasar-like continuum spectral energy distribution (SED, e.g. Mathews & Ferland 1987, Elvis et al. 1994) tends to develop thermal instabilities, due to line cooling, that allow gas in two or more phases to co-exist in pressure equilibrium at different temperatures but, naturally, at quite different densities and ionization parameters[2]. Chakravorty et al. (2008, 2009, 2012) updated the Krolik et al. (1981) calculations with recent atomic physics parameters and covered a wide variety of possible quasar SED shapes. They found that multi-phase media form only under certain conditions (Chakravorty et al. 2012).

One condition is that the BELR gas have super-Solar metallicity ($Z/Z_\odot > 2$). Super-Solar metallicity is a normal occurrence in quasars (Hamann & Ferland 1999, Hamann et al., 2007). These authors show that the NV/CIV BEL emission line ratio scales roughly linearly in log-log space with $Z/Z_\odot$, with values of log(NV/CIV) > -1 implying $Z/Z_\odot > 1$. Log(NV/CIV) is larger (~ -0.3 – +0.3) in high luminosity quasars [$\nu L_\nu \geq 10^{13}$ $L_\odot$(1450Å)] implying $Z/Z_\odot$ ~10 – 20, but is still mostly > -1 for luminosities down to $\nu L_\nu \geq 10^{10}$ $L_\odot$(1450Å), implying $Z/Z_\odot$ ~2 – 5 (Hamann & Ferland 1999). Supporting high $Z/Z_\odot$ values in quasars, Groves et al. (2006) found that the narrow line regions had $Z/Z_\odot$ ~2 – 4, and Fields et al. (2007) find evidence for $Z/Z_\odot > 2$ in a WA. Galaxy bulges (e.g. the Milky Way, Johnson et al. 2014) typically have $Z/Z_\odot > 1$.

Another constraint is that the quasar continuum SED should lie within certain bounds (Chakravorty et al., 2009, 2012). In fact, the intrinsic SED shapes of quasars are remarkably constant once host galaxy contributions and reddening effects are removed (e.g. Hao et al. 2013, Lusso and Risaliti 2016). There is a weak dependence of the ratio of X-ray to UV luminosity on UV luminosity, with little intrinsic scatter (Risaliti and Lusso 2015, Lusso and Risaliti 2016).

Given these two results a cool phase is commonly expected to form in quasar BELRs.

In almost all cases the breadth of the multi-phase region in pressure space is no more than a factor 2-3, and the low temperature phase spans only a factor ~3 in temperature (Chakravorty et al. 2009), so that the condensations in a given AGN will then effectively have constant density.

Denser gas has a larger effective optical depth and this reduces the force multiplier over electron scattering, $\mathcal{M}$, roughly to the 0.8-0.9 power of density for quasars (MCGV).

---

[2] The ionization parameter, U, is defined as the ratio of ionizing photon density to the electron density: U = (Q/4π$R^2$c)/$n_e$. Here Q is the number of ionizing photons per second emitted by the quasar continuum source, R is the distance of the BEL gas from that source, and c is the speed of light (Frank, King & Raine 2002).

Hence for phases in pressure equilibrium, the factor 100 higher density of the $10^4$ K cool phase over the $10^6$ K warm phase, reduces $\mathcal{M}$ from the typical values of 10 – 30 (Risaliti & Elvis 2010) to the minimum, electron scattering only value, of unity. This cooler gas experiences only Thompson cross-section momentum transfer from the radiation field and so is effectively in a sub-Eddington state and cannot escape. As Marconi et al. (2008) showed clouds with $N_H > 10^{23}$ cm$^{-2}$ (section 3.4), experience virtually no radial acceleration.

*2.2 Observational*

Cool (~$10^4$ K) gas is seen almost universally in quasars in three forms: (1) BELs are a defining feature of quasars (section 1). (2) WAs are seen in more than half of all quasars (Piconcelli et al. 2005, Laha et al. 2014) and a cool (~$10^4$ K) low ionization phase (LIP) in pressure equilibrium with the warmer (~$10^5$ K – $10^6$ K) phase is also normal[3]. [But see Steenbrugge et al. (2003, 2005) and Blustin et al. 2004) for alternative interpretations.] Krongold et al. (2005) note that the WA must be heavily clumped, but there are few measurements constraining this clumping. (3) X-ray eclipsers are harder to find because of the intensive monitoring required to do so, but Torricelli-Ciamponi et al. (2014) found that at least ~1/3 (15/42) of AGN showed X-ray eclipses. Note that any flattening of the relevant region will make the two types of absorbers observationally less than universal, even if they are always present (see section 2.3).

The temperatures, densities, ionization parameters and distances from the central continuum for all three signatures of cool gas in quasars are all consistent (Table 1, Figure 2):

1. Temperatures for the BELR gas are ~1- 4 ×$10^4$ K (Osterbrock & Ferland 2006, Popović 2003). The low ionization WA phase is at $(2.5 – 3.7) \times 10^4$ K (Krongold et al. 2003, 2005a, 2007, 2009, Andrade-Velasquez et al. 2010). X-ray eclipsing clouds must have T ~< $10^5$ K in order not to highly ionize oxygen and make the gas transparent below ~0.5 keV.
2. Density estimates for BELR gas lie in the range $n_e = 10^9 – 10^{11}$ cm$^{-3}$ (Ferland et al., 1992, Osterbrock & Ferland 2006). The few measured recombination and ionization times for warm absorbers (Krongold et al. 2005, 2007) give low ionization phase (LIP) densities of $>10^4$ cm$^{-3}$ and $>10^8$ cm$^{-3}$ for NGC3783 and NGC4051 respectively, and $>10^4$ cm$^{-3}$ or MR2251-179 (Reeves et al., 2013). This measurement is a challenging one for existing X-ray spectrometers, however. X-ray eclipses also require $n_e$ ~$10^{10} – 10^{11}$ cm$^{-2}$ (Risaliti et al. 2009ab, Maiolino et al. 2010).
3. Ionization parameters for BELR gas lie in the range *log U* = -1.5 for CIV to *log U* = 1 for OVI and *log U* < 1 for the Balmer lines (Korista et al. 1997). The WA multi-phase solutions have *log U* ~ -1 to *log U* ~ 0 for the LIP (Krongold et al. 2003, 2005b, 2007 Andrade-Velazquez et al. 2010). The X-ray eclipsing clouds have low, but poorly constrained ionization (Risaliti et al. 2007).

---

[3] Mainly in work by Krongold and collaborators [NCG 3783 (Krongold et al. 2003; Netzer et al. 2003), NCG 985 (Krongold et al. 2005ab , 2009), NGC 4051 (Krongold et al. 2007), Mrk 279 (Fields et al. 2007), NGC5548 (Andráde-Velasquez et al. 2010, Krongold et al. 2010)]; see also Holczer, Behar & Kaspi (2007) and Laha et al. (2014) who find a gap in ionization parameter values in the WA gas.

4. Distances of the BELR from the continuum span a factor 10 (Peterson & Wandel 1999) with Hβ being at ~ a few 1000 Schwarzschild radii, $r_g$. The X-ray eclipsers are at similar distances (Elvis et al. 2004, Puccetti et al. 2007, Risaliti et al. 2005, 2007, 2009a,b, 2011). WAs can be distant (Crenshaw and Kraemer 2012, Arav et al. 2013), but others are close in at a few 1000 $r_g$ (Krongold et al. 2005b, 2007).

It is then reasonable to treat all three observed phenomena as different aspects of the same gas clouds.

Table 1: Physical parameters for warm components in AGN (see Section 2.2)

| Parameter | BELR | WA Low Ionization Phase | X-ray eclipsing clouds | Overlap Region |
|---|---|---|---|---|
| Temperature, T (K) | (1-2) x $10^4$ | *few* x $10^4$ | <$10^5$ | **(1-4)×$10^4$** |
| Density, $n_e$ (cm$^{-3}$) | $10^9$ - $10^{11}$ | **>$10^4$,** > $10^8$ | $10^{10}$ - $10^{11}$ | $10^{10}$ - $10^{11}$ |
| Ionization parameter, U. | $10^{-1.5}$ - $10^1$ | $10^{-1}$ – 0 | <100 | 0.1 – 0 |
| Distance from continuum, R ($r_g$) | $10^3$ – $10^4$ | Few 1000[a] | Few 1000 | 1000 – 10000 |

a. but see section 2.2 for references to different distances.

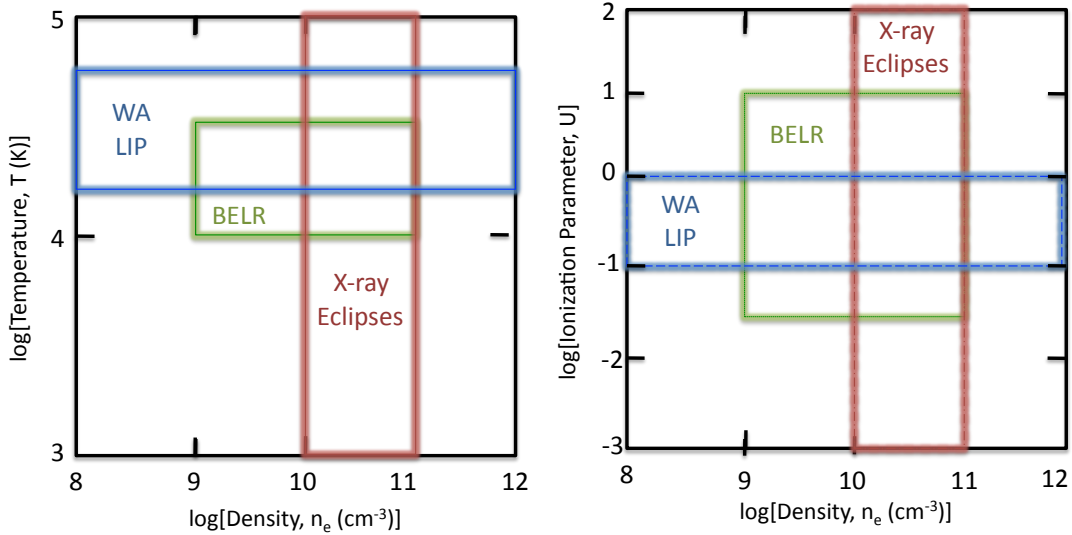

**Figure 2.** Typical values in: (*left*) density-temperature plane and (*right*) density-ionization parameter plane, for BELR (green), Warm Absorber (WA) low ionization phase (blue) and X-ray eclipser (red) gas in AGNs.

*2.3 Inflows and Winds: Model Starting Points*

The model presented here was spurred by velocity resolved reverberation mapping (VRRM) results that show evidence for inflows at speeds of ~(1000– 2000) km s$^{-1}$ in the BELR (Bentz et al. 2010, Grier et al. 2013). This unexpected result is hard to escape as zero or short lag times must be associated with gas along or near to our line of sight to the central continuum source. Hence redshifts (relative to the quasar host galaxy) at short lags are unambiguously due to inflow[4].

The kinematic models that Pancoast et al. (2014) fitted to the VRRM data find that many of the BEL clouds are moving on slightly elliptical orbits. This surprising feature is hard to incorporate in most BELR models (but see Gaskell 2009), and is the starting point of this model.

Pancoast et al. (2014) also found that, in 3/5 cases, an optically thick screen in the plane of the distribution that hides the far side BEL emission from us. This last result suggests that the BELR is associated with the accretion disk and may be related to an accretion disk wind (e.g. Shields 1977, Murray and Chiang 1998, Elvis 2000, Waters et al., 2016). We see such winds in quasar WAs, as blueshifted absorption lines in both the X-ray and ultraviolet (UV). This association of the BELR and the accretion disk wind is the starting point of this model. In this model the outflow is the inflow, or at least its source.

Accretion disk winds in quasars can be accelerated by several means: thermal expansion, radiation pressure, both electron scattering and line driving, and magnetic slingshot (Ohsuga and Mineshige 2014). Line driving by the strong quasar ultraviolet (UV) continuum (e.g. MCGV) is especially appealing as it uses both the strongest part of the quasar output and the large force multiplier, $\mathcal{M}$, that radiation line driving can provide over the Thomson electron scattering cross-section, as found in O-star winds (Castor, Abbott and Klein 1975, Lamers & Cassinelli 1999.) Bowler et al. (2014) find evidence for CIV line locking in 2/3 of SDSS quasars with absorption lines, which argues strongly for line-driven outflows.

For quasar SEDs $\mathcal{M} \sim 10 - 30$ (MCGV, Risaliti and Elvis 2010). Typical quasars have Eddington ratios, $L/L_{Edd} \sim 0.1$ (Kollmeier et al., 2006, Steinhardt & Elvis 2010). As the standard Eddington ratio is defined using the Thomson electron scattering cross-section alone, quasars with values of $\mathcal{M} > \sim 10$ will exceed their effective Eddington limit for line driving. Hence escaping disk winds are expected to be a normal feature of quasars, so long as the X-ray flux is not too strong relative to the driving UV continuum, which will over-ionize the gas (MVGC, Murray & Chiang 1995, Proga 2003, Risaliti & Elvis 2010). Line driving will be assumed hereafter in this paper.

Line driven wind models for quasars tend to have flattened BELRs, ranging from almost disk-hugging (Murray and Chiang 1995) through intermediate angles (Elvis 2000, Risaliti and Elvis 2010) to more volume filling but with a hollow core (Proga 2003). Encouragingly, the models of Pancoast et al. (2014) find quite flattened BEL cloud distributions. However, the quasar rain model presented here is not yet detailed enough to address this flattening.

---

[4] Optically thick BEL clouds will primarily radiate their resonance lines back toward the ionizing continuum source. This is a well-known result in AGN physics (Netzer 1975, Osterbrock and Ferland 2006). The models of Pancoast et al. (2014), which deliberately do not include radiative transfer physics, show that most of the flux from the BELs comes from clouds on the far side of the nucleus, corroborating this result. The small flux from the unilluminated backs of the near side clouds must produce the infalling zero-lag redshifted emission (Bentz et al. 2010, Grier et al. 2013).

# 3. TIMESCALES FOR CONDENSATIONS IN AGN OUTFLOWS

For the quasar rain picture to work the condensation time for cool clouds to form out of the warm outflow, $\tau_{cool}$, must be less than the escape time, $\tau_{esc}$, which is the time it takes for the wind to accelerate to the escape velocity, $v_{esc}$. The clouds must also not be destroyed so quickly that they are not seen. Below we show that the timescales involved all have values that allow the conditions for quasar rain to be satisfied. Table 2 summarizes the timescales derived in the following sections.

The ranges of relevant parameters for quasars are given in Table 2. These span the most typical values encountered in quasars. The cool cloud density range comes from the overlap region in figure 2. The values for the timescales calculated for the baseline set of parameters and their ranges of values in Table 2 are shown in Table 3.

Table 2: Baseline parameter values and ranges.

| Parameter | Symbol | Minimum value | Baseline value | Maximum value | Units |
|---|---|---|---|---|---|
| Initial cloud radius | $r_i$ | $3 \times 10^{12}$ | $1 \times 10^{13}$ | $2 \times 10^{14}$ | cm |
| Initial cloud temperature | $T_{i,6}$ | $5 \times 10^5$ | $1 \times 10^6$ | $1 \times 10^6$ | K |
| Initial WA radial velocity | v | 200 | 1000 | 2000 | km s$^{-1}$ |
| Black hole mass | M | $10^7$ | $10^8$ | $10^9$ | $M_\odot$ |
| Metallicity | Z | 2 | 3 | 10 | $Z_\odot$ |
| Cool cloud density | $N_e$ | $10^{10}$ | $10^{10}$ | $10^{11}$ | cm$^{-3}$ |
| BELR distance | R | 1000 | 3000 | 10,000 | $r_g$ |
| L/L$_{Eddington}$ | $\lambda$ | 0.01 | 0.1 | 1 | -- |
| Mach number | $M$ | 10 | 20 | 100 | -- |
| Cloud density contrast | $\chi$ | 100 | 100 | 100 | -- |

Table 3: Timescales in the Quasar Rain model

| Timescale | | Equation[a] | Baseline[b] | Range[c] | Ref[d] |
|---|---|---|---|---|---|
| Dynamical | $\tau_{dyn}$ | $(R^3/GM)^{1/2} = P_{orb}/2\pi = 1.4\ R_{1000}^{3/2}\ M_8$ *yr* | 7 yr | 0.15 - 450 yr | 1 |
| Escape | $\tau_{esc}$ | $v_{esc}/g = (v_{esc}/v_{Kep}) \cdot \tau_{dyn} = 1.4\ \tau_{dyn}$ s | 10 yr | 0.20 - 650 yr | 2 |
| Apocenter | $\tau_{apo}$ | $v(radial)/g = 75 \cdot v_{1000} \cdot R_{1000}^2\ M_8$ d | 2 yr | 1.5d – 400 yr | |
| Cooling time | $\tau_{cool}$ | $2 \times 10^4\ [\Lambda(T)/\Lambda_b(T)]^{-1}\ (Z/Z_\odot)^{-1}\ T_{i,6}^{1/2}\ n_{10}^{-1}$ s | 5 hr | 400s - 8 hr | 3 |
| Sound crossing | $\tau_s$ | $2r/c_s \sim 1.3 \times 10^6\ T_{i,6}^{-1/2}\ r_{i,13}$ s | 15 d | 3d – 1.2 yr | 4 |
| Cloud crossing | $\tau_{ic}$ | $\tau_s/M = 2r_c/Mc_s$ s | 1 d | 0.03 - 45 d | 5 |
| Cloud crushing | $\tau_{cc}$ | $\chi^{1/2}\tau_{ic} = 2\chi^{1/2}r_c/Mc_s$ s | 8 d | 0.3 – 1.2 yr | 5 |
| X-ray | $\tau_X$ | $M_\odot^2/2.5 \times 10^6 \cdot L_{bol,44} = 0.32 M_8 (L/L_{Edd})^{-1}$ d | 3 d | 0.03 – 300 d | 6 |

a. Where:
  $c_s = (\gamma kT/\mu m_H)^{1/2} = 150\, T_6^{1/2}$ km s$^{-1}$;
  $g$ is the local acceleration due to gravity, $GM/R^2$;
  G is the gravitational constant;
  k is Boltzmann's constant = $1.38 \times 10^{-16}$ erg K$^{-1}$;
  $L/L_{Edd}$ is the Eddington ratio;
  $L_{bol,44}$ is the ultraviolet bolometric luminosity in units of $10^{44}$ erg s$^{-1}$;
  *M* is the Mach number;
  M is the mass of the black hole in solar masses;
  $M_8$ is M in units of $10^8$ solar masses;
  $m_H$ is the mass of the hydrogen atom = $1.67 \times 10^{-24}$ g;
  $P_{orb}$ is the orbital period in s;
  R is the distance from the central black hole in cm;
  $R_{1000}$ = is R in units of 1000 Schwarzschild radii, $r_g = 2GM/c^2$;
  $r_{i,13}$ is $r_i$ the initial radius of a condensing cloud in units of $10^{13}$ cm;
  $r_c$ is the radius of the condensed cloud, = $r_i \cdot \chi^{-1/3}$, for a density ratio of $\chi$; i.e. 0.22 $r_i$ for a density ratio of 100;
  $T_{i,6}$ is the initial temperature of the wind in units of $10^6$ K;
  $v_{1000}$ = initial radial WA velocity in units of 1000 km s$^{-1}$;
  $v_{esc} = (2GM/R)^{1/2}$ is the escape velocity from radius R = 9500 $R_{1000}^{-1/2}$ km s$^{-1}$ = $\sqrt{2} \times 6700\, R_{1000}^{-1/2}$ km s$^{-1}$;
  $Z/Z_\odot$ is gas metallicity relative to solar (section 2.1);
  $\Lambda(T)$ is the cooling coefficient (erg s$^{-1}$ cm$^3$) ;
  $\Lambda_b(T)$ is the cooling coefficient for bremsstrahlung;
  $\Lambda(T)/\Lambda_b(T)$ is the factor increase in the cooling coefficient in a thermal plasma due to line cooling over bremsstrahlung at solar metallicity, which has values of ~35 for T = $10^5$ - $10^6$ K, ~100 for T = $10^{4.5}$ K, and peaks at ~500 for T = $10^{5-5.5}$ K;
  $\gamma$ is the ideal gas adiabatic index = 5/3;
  $\mu$ is the mean molecular weight of the gas (~0.6); and
  $\chi$ is the ratio of the cloud density to the ambient medium density.

b. Using the baseline values from Table 2.

c. Using the minimum and maximum values from Table 2.

d. References: (1) Frank, King & Raine (2002); (2) MCVG; (3) Tucker (1975), Gehrels & Williams (1993); (4) Klein, McKee & Colella (1994); (5) Klein, McKee & Colella (1994), Patnaude & Fesen (2014); (6) McHardy et al. (2006), Fender et al. (2007).

### *3.1 Wind Escape versus Cloud Condensation Timescales*

   The time to reach escape velocity for the WA wind, $\tau_{esc}$, can be approximated as the wind velocity over the local gravity, **$v_{esc}$**/$g$, where $g = v_{Kep}^2/R$, and $v_{Kep}$ is the Keplerian orbital velocity at radius R. Hence $\tau_{esc}$ = (**$v_{esc}$**/$v_{Kep}$) . $\tau_{dyn}$, the dynamical timescale[5], so $\tau_{esc}$ = $\sqrt{2}\, \tau_{dyn}$. As the wind material is ejected from the accretion disk it must reach $v_{esc}$ within about half an orbit or else impact the disk again, which is assured by $\tau_{esc}$. Longer values of $\tau_{esc}$ are found for more distant BELRs at $R_{1000}$ = 10 around higher mass black holes ($M_8$ = 10).

---

[5] I thank Norm Murray for this concept.

The cloud cooling time ($\tau_{cool}$, Table 3) for a thermal plasma at these temperatures is short because it is dominated by strong line cooling. At T = $10^6$ K the cooling coefficient, $\Lambda$, is ~$10^{-22}$ erg s$^{-1}$ cm$^3$, 30 - 40 times larger than for pure bremsstrahlung ($\Lambda_b$, Gehrels and Williams 1993), for which the cooling time is $\tau_{cool,b}$ ~ $2\times10^4$ $T_6^{½}$ $(Z/Z_\odot)^{-1}$ $n_{10}^{-1}$ s (e.g. Tucker 1975). Given cloud densities of $10^{10}$-$10^{11}$ cm$^{-3}$ (Table 2, Figure 2), the initial warm medium must have had a density ~$10^8$-$10^9$ cm$^{-3}$. So $\tau_{cool}$ ~ 400 s (for $Z/Z_\odot$ = 10, $T_6$ = 1, and $n_8$ = 10) and ~5.9 hours (for $Z/Z_\odot$ = 2, $T_6$, $n_8$), as in Table 3. The cooling coefficient is over 10 times larger at $10^{5-5.5}$ K, so the cooling time will be dominated by the initial drop from $10^6$ K to $10^{5.5}$ K. This calculation is only a rough guide to $\tau_{cool}$ as the WA is photoionized, not collisional. A more careful treatment would be valuable.

Much more restrictively, the initially warm (~$10^6$ K) gas cannot collapse on timescales shorter than the sound crossing time, $\tau_s$. From X-ray eclipses, we know that the condensed clouds have sizes ~$6 \times 10^{11}$ - $5 \times 10^{13}$ cm (Maiolino et al., 2010, Risaliti et al 2009a), and would have had an initial size, $r_i$ ~ $3 \times 10^{12}$ - $2 \times 10^{14}$ cm in the warm high ionization phase, so $\tau_s$ is of order weeks (Table 3).

As $\tau_s \ll \tau_{esc}$ condensed cool clouds will generally be present in quasars.

Clouds will reach their apocenter, and so zero radial velocity, on a timescale $\tau_{apo}$ = v(radial)/g = 75.$v_{1000}$.$R_{1000}^2$ $M_8$ days, spanning a wide range from 1.5 d to 400 years. For clouds at the inner edge of the BELR, $R_{1000}$ = 1, this is 75 days, while at the outer edge, $R_{1000}$ = 10, it becomes 20 years. After this time the clouds will be infalling and will fall back on a dynamical timescale, $\tau_{dyn}$ ~0.2 - 500 year (Table 2), "raining out" of the WA medium.

As $\tau_{cc} < \tau_{apo}$ there is the possibility that clouds are destroyed on their outbound leg. At first $\tau_{cc}$ will be long but, as the cloud slows, $M$ increases, reducing $\tau_{cc}$. At apocenter a cool cloud at $10^{4.5}$ K has $c_s$ ~25 km s$^{-1}$, and so is at Mach 40 in a 1000 km s$^{-1}$ wind and will be crushed in a few days. However, the velocity law within the wind in MCGV is mostly linear with radius, so small, low $N_H$, clouds that form rapidly in the outflowing wind will form at smaller radii, when the wind is slower. They will then have smaller Mach numbers and will last 10 – 100 times longer than the $10^{23}$ cm$^{-2}$ clouds seen in X-ray eclipses to date. (And they will have wider Mach cones – see Section 3.6.) This is likely an important effect deserving further investigation.

### *3.2 Cool Cloud Formation Mechanism*

Condensing gas into cooler clouds is not enough, as they may fall anywhere on the thermal equilibrium curve in pressure-temperature space. In order to form a multi-phase medium in pressure equilibrium the gas must find its way to stable locations. Sound waves originating at the base of the wind in the disk photosphere, as in O-star winds, could amplify, perturbing gas off the thermal equilibrium curve (Lucy & White 1980) and initiating a collapse. Another mechanism that I propose here relies on variability of the continuum source shape. This mechanism is unique to quasars because of their strong, rapidly variable, X-ray continuum.

The value of the radiation line driving force multiplier, $\mathcal{M}$, is non-linearly sensitive to the continuum intensity and shape (MCGV). In particular X-rays can overionize the gas, reducing ℳ dramatically (MCGV). Stochastic variations in the X-ray/UV flux ratio will change ℳ, and this will necessarily lead to quasi-random compression and rarefaction throughout the wind. (These sound waves may also amplify and shock, as in O-stars, MacGregor et al. 1979.)

Material then moves off the thermal equilibrium line. The gas heats or cools until it returns to this curve. Gas in stable regions of the curve will remain there, unless the perturbations are large. Gas in other regions will likely return to new points on the curve. If the X-ray variations are on timescales $\tau_X \ll \tau_{esc}$ then many perturbations will occur before the gas reaches $v_{esc}$. Gas will then aggregate on the stable branches of the thermal equilibrium curve, including the low temperature, high-density phase in the multi-phase region.

Observationally, $\tau_X \ll \tau_{esc}$ for most parameter values (Table **3**). The central continuum that ionizes and heats the winds in quasars is irregularly variable, in the UV by modest factors on weeks to years timescales (Vanden Berk et al. 2004), and by similar or greater factors in X-rays on an hours to months timescales (McHardy et al. 2006, Fender et al. 2007). Low mass black holes accreting at high rates vary quickest (Table 2). The UV and X-ray fluxes are correlated on these short timescales when the longer-term trends are removed, but with smaller UV amplitude changes (McHardy et al. 2014 and references therein). The X-ray/UV flux ratio then changes continually by factors of a few, changing ℳ. A rapidly changing ℳ should ensure that the stable branches on the thermal equilibrium curve are all occupied, including both warm ($10^6$ K) and cool ($10^4$ K) phases. High mass, low accretion rate quasars have $\tau_X \sim \tau_{esc}$ and will have difficulty populating the stable regions on the thermal equilibrium curve.

The thermal stability curves, on which the multi-component phases in the wind relies, are derived assuming an equilibrium case (Krolik et al. 1981, Chakravorty et al. 2010). However this may not be the case, as the cooling time that determines the equilibrium time is comparable to the X-ray variability timescale (Table 3). A more detailed treatment of this process could be worthwhile.

### *3.3 Raindrop Destruction versus Infall Timescales*

The condensed cool clouds cease accelerating (Section 2.1) and so will start to feel a pressure from the accelerating wind that surrounds them. The sound speed of the cool gas, $c_s$ = 25 km s$^{-1}$ at T = 3 ×10$^4$ K (Table 3). The sound speed of the surrounding ~10$^6$ K medium will be ~100 km s$^{-1}$, well below the wind speed. Hence the condensed clouds will rapidly become supersonic relative to the warm wind. Mach numbers, $M$, from 10 to ~100 will be reached, using the typical observed WA velocities of 1000 km s$^{-1}$ and $v_{esc}$ respectively.

The effects of supersonic flows on BELR clouds have been studied for decades (e.g. Krolik 1977, Blumenthal and Mathews 1979, Mathews and Ferland 1987). More generally, the destruction of cool clouds exposed to a hot supersonic flow has been widely studied in a variety of astrophysical contexts: in O-star winds (Klein, McKee & Colella 1994), supernovae (e.g. Patnaude & Fesen 2014) and quasar feedback on host galaxy molecular clouds (e.g. Hopkins & Elvis 2010).

The cloud will start to be crushed when it becomes supersonic. As the sound speed in the cloud is only ~25 km s$^{-1}$, the wind only has to accelerate by ~0.2 – 0.3% of v$_{esc}$ after the cloud forms before this is satisfied. Naively taking 0.2 – 0.3% τ$_{esc}$ as the time to reach that velocity, cloud crushing begins after 4 hours – 1.5 years. This is comparable to the cloud crushing time we find below.

At $M \sim 10 - 100$, Rayleigh-Taylor instabilities will disrupt the cloud, allowing Kelvin-Helmholtz instabilities to ablate gas from the side surfaces of the clouds. [Although Vietri, Ferrara and Miniati (1997) suggest that radiative losses will stabilize the clouds somewhat.] Estimates of the destruction time for the BELR raindrops under these conditions imply short lifetimes. Initially this result was taken to mean that there was most likely no hot confining medium (e.g. Mathews and Ferland 1987). The idea that BELR clouds are transient structures that are continually replenished allows the presence of a confining medium. This idea has since gained support from the discovery of winds in quasars (e.g. Elvis 2000). The quasar rain model assumes a transient/replenished scenario for the BELR clouds.

The X-ray eclipse observations give cool cloud radii of ~$6 \times 10^{11}$ - $5 \times 10^{13}$ cm (Maiolino et al. 2010, Risaliti et al. 2009a) in reasonable agreement with the estimates in Section 3.4. The sound crossing time $\tau_s \sim 3$ - 450 days, but a supersonic shock wave crosses the cloud $M$ times faster, $\tau_{ic} = \tau_s/M$. For representative Mach numbers, $M = 10$ - 100, a shock can cross a cloud in **$\tau_{ic}$ = 0.03 - 45 days**.

A longer, and more relevant timescale is the cloud crushing time, $\tau_{cc} = \chi^{1/2}\tau_{ic}$, where $\chi$ is the ratio of cloud to ambient medium densities (Klein, McKee & Colella 1994, Patnaude & Fesen 2014). A cool ($10^4$ K) cloud in pressure equilibrium with a warm ($10^6$ K) gas of density **n$_e$** must have a density of 100 **n$_e$**, so $\chi \sim 100$ for the BELR. Hence, $\tau_{cc} \sim$ 0.3 - 450 days. This is far shorter than the dynamical infall time for the clouds, $\tau_{dyn}$.

### *3.4 Column Densities for the Raindrops*

We can use these cloud formation and destruction timescales to set some bounds on the range of column densities, N$_H$, of the condensed clouds. An upper limit can be found by realizing that if the cloud formation time, $\tau_s$, becomes larger than the escape time, $\tau_{esc}$, then the cloud will escape. Using Table 3 and a factor 4.6 between initial and final cloud radii (for a factor 100 decrease in volume) gives a cool cloud radius, r$_c$ < $10^{14}$ R$_{1000}^{3/2}$ M$_8$ T$_6^{1/2}$ cm. For the range of values in Table 2, this gives r$_c$ < $7.5 \times 10^{12}$ – $3.5 \times 10^{16}$ cm. Since N$_H$ = 2r$_c$·**n$_e$**, N$_H$ < $1.5 \times 10^{(23-24)}$ – $2.4 \times 10^{(26-27)}$ cm$^{-2}$. The two ranges are for low and high BELR densities (Table 2). The lower range of values are appropriate for most of the reverberation mapped quasars, which have masses of M$_8$ = 0.1 (Peterson et al., 2004), and are in agreement with observational values (e.g. Risaliti et al. 2009a, 2011, Netzer, 2013). The highly Compton thick ($\tau$ > 100) values would apply at large black hole masses, M$_8$ = 10, but may be unphysical. It would be worth investigating if such clouds produce observable signatures.

Large clouds may be characteristic of the narrow emission line region in quasars (e.g. Fischer et al., 2011, 2013). The ability of large clouds to reach v$_{esc}$ before condensing may explain this observation.

In addition, to condense out of, and be in pressure balance with, a hotter medium, BELR clouds cannot be so large that they leave no surrounding medium. We can say roughly that they cannot take up more than a ~10° arc around the orbit they begin at, i.e. their radius cannot be more than of order a tenth of their distance from the ionizing source, R. So $r_c < 0.1$ R. This condition gives a more reasonable upper bound than the above cooling time argument. At $R_{1000} = 1$ and $M_8 = 1$, $r_c \sim\!< 3 \times 10^{15}$ cm, so $N_H \sim\!< 6 \times 10^{(25-26)}$ cm$^{-2}$, depending on the assumed BELR density. This limit is more restrictive for smaller black holes or for BELR clouds radiating higher ionization lines (e.g. CIV, HeII) that lie at 5 – 10 times smaller distances from the continuum than Hβ (e.g. Grier et al., 2012).

Together the two limits give $10^{22} < N_H < 10^{25}$ ($n_e = 10^{10}$ cm$^{-3}$) and $10^{23} < N_H < 10^{26}$ ($n_e = 10^{11}$ cm$^{-3}$). Compton thin clouds it seems may not escape, while Compton thick clouds may. This dichotomy may have observable effects.

### *3.5 Orbits of the Raindrops*

The sub-escape velocity BELR clouds, or "raindrops", will move on elliptical orbits, as they began on circular Keplerian orbits in the accretion disk and are then given an effectively impulsive outward radial kick. Pancoast et al. (2014) find that a large fraction (20% - 60%) of BELR clouds lie on elliptical orbits in four of the five AGN that they investigated. In the quasar rain model the ellipticity of their orbits are typically quite small. It can be shown that[6], given a radial kick $\Delta v$ from an initial orbital velocity $v_{orb}$, the fractional increase of the semi-major axis *a* over the original semi-major axis $a_0$ is $a/a_0 = 1/(1 - (\Delta v/v_{orb})^2)$, and the eccentricity, $e = (\Delta v/v_{orb})^2$. For $\Delta v \sim 1000$ km s$^{-1}$, based on observed WA velocities, and at $R_{1000} = 1$ $v_{orb}(1000) = 6700$ km s$^{-1}$. So $e = 0.02$ and $a/a_0 = 1.02$. Instead at $R_{1000} = 10$, $e = 0.22$ and $a/a_0 = 1.29$. (Note, by the way, that a radial kick is an inefficient way to eject clouds[7]).

The inclination of the cloud orbits to the accretion disk will be that of the wind at the point where the clouds form. Clouds can then always create X-ray eclipses in any of the >50% of quasars in which a WA is seen. X-ray eclipses should not be seen in quasars without a WA. Interestingly the best studied X-ray eclipse source, NGC1365 (Section 3.6) has recently been found to have a WA when observed in a rare low obscuration state (Braito et al., 2014), consistent with this prediction.

A few notes are of interest: (1) faster moving clouds on more elliptical orbits will be destroyed proportionately faster, so all clouds fall a similar distance; and (2) when the ellipticity of the cloud orbits is small the ambient pressure of the high ionization phase of the warm absorber medium will decrease only by a few percent on the outbound part of the elliptical cloud orbit. As a result the raining clouds will remain pressure confined and their ionization parameter will not be changed substantially.

As the infalling clouds do not fall far their infall velocities will be limited and comparable to their outflow speed at the time of their formation. Their motion will be dominated by the rotational velocity the wind material had on leaving the accretion disk. They have only the angular momentum from their point of origin.

---

[6] I thank J.C. McDowell for this derivation.

[7] A velocity kick >$v_{orb}$ ($\Delta v/v_{orb} > 1$) is needed to produce an escaping cloud as, although $v_{esc} = \sqrt{2}$ $v_{orb}$, the two velocity vectors are perpendicular to one another. The most efficient method is to apply a $\Delta v$ tangential to the orbit in the direction of motion, which need only be ($\sqrt{2} - 1$)$v_{orb}$ to achieve $v_{esc}$. This is the technique used for interplanetary spacecraft.

### *3.6 Ablating Raindrops as Cometary X-ray Eclipsers*

Asymmetric X-ray eclipse profiles with a rapid (< 2 ksec) onset followed by a more gradual (~50 ksec) recovery, were found in two events in NGC 1365 (Maiolino et al. 2010). These imply a "cometary" head-tail structure for the eclipsing clouds, which was puzzling. The quasar rain picture of infalling raindrops with large transverse motions provides a natural explanation for these cometary X-ray eclipses.

In particular the cloud destruction parameter values are all in accord with the observations:

(1) The ablating raindrops will produce comet-tail-like structures behind the direction of motion of the clouds.
(2) At Mach numbers, $M \sim 10 - 100$ for the raindrops the opening angles of the clouds' Mach cones will be $1/10 - 1/100$ radians, $\sim 6 - 0.6$ degrees. Very comparable opening angles (< 1.2, 4.8 deg) are measured for the two asymmetric X-ray absorbing eclipses in NGC1365.
(3) The small ellipticities of the cloud orbits (section 3.5) imply that the tails must be largely transverse when they cross our line of sight. They would then reproduce the highly asymmetric X-ray eclipses.
(4) The predicted lifetime of the ablating clouds is ~months (Table 3), in agreement with the observed lifetimes (Maiolino et al. 2010).

## 4. DISCUSSION, CONSEQUENCES OF THE MODEL

The basic physics of quasar rain is commonly encountered in astronomy. The idea of condensations forming in a warm medium has been proposed as a means of accreting material out of the cool cores of clusters of galaxies (Voit & Donahue 2014, Gaspari, Brighenti & Temi 2015). Producing elliptical orbits via a radiative kick has also been proposed as a way to induce "positive feedback" onto supermassive black holes in order to promote their growth by stimulating the accretion of molecular clouds (Dehnen & King 2013). Cometary head-tail structures are seen in the HI emission of High Velocity Clouds in the halo of the Milky Way (e.g. Heitsch and Putman 2009). The physics of the model are suitably robust; they will occur for most supermassive black holes accreting at reasonable fractions of the Eddington rate.

Here I explore further the consequences of quasar rain, list some tests of the model and then discuss some challenges to the model.

### *4.1 Origin of BELR Cloud material*

The short raindrop lifetimes mean that the supply of BELR clouds must be continually renewed. Maiolino et al. (2010) suggested that the only nearby source was the accretion disk. In the quasar rain scenario this picture is partly modified in that the immediate source of the BELR clouds is the WA outflow. This outflow itself though originates in the accretion disk, which is then, indirectly, the source of the BELR material.

## 4.2 Origin of Constant Density BELR Clouds

There is a puzzling scaling of BELR radius with continuum luminosity, $L_{bol}$: r(BELR)=$const.L_{bol}^{1/2}$ (Bentz et al. 2006, 2013). This scaling suggests that all BELR clouds have the same density, so that they will have constant U at radii proportional to $L_{bol}^{1/2}$. If BELR clouds form in a 2-phase equilibrium then this relation follows naturally.

The small range of pressures and temperatures at which stable cool condensations can form a second phase in pressure equilibrium with a hotter medium (section 2.1, Chakravorty et al. 2009) implies that these clouds will all have similar densities. The observed Bentz et al. (2006, 2013) relation then follows naturally within a given quasar as it varies in ionizing flux (Peterson et al. 1999). Since the spectral energy distributions SEDs of quasars are reasonably constant (e.g. Hao et al. 2013, Lusso and Risaliti 2016) $L_{bol}$ becomes the only variable and the observed BELR radius scaling with $L_{bol}^{1/2}$ is expected in a population of quasars.

Constant density clouds are also implied within an individual BELR. It is well established that the BELR is radially stratified by ionization state with higher ionization clouds being closer in, based on reverberation mapping data (Peterson and Wandel, 1999). Qualitatively constant density clouds would produce this stratification. The run of ionization parameter with distance is not well known, though, so we cannot tell if the BELR clouds have truly constant density with radius.

## 4.3 $N_H$ distribution and X-ray Eclipses

The ~5 eV resolution of current X-ray calorimeters (Mitsuda et al. 2014) gives a resolving power of ~1000 (~3000 km s$^{-1}$) at Fe-K. This high resolution allows weak absorption lines to be detected, allowing probes of lower column density eclipsing clouds. Unfortunately, following the demise of *Hitomi*[8], it will be some years before an X-ray calorimeter is flown that can investigate this prediction.

The small range of pressures at which a low ionization phase exists for a given input spectrum (Sec.2.1, Chakravorty et al., 2009) implies that the lower $N_H$ clouds in a given quasar should tend to be smaller and so have lower covering factors, $f_c$, for the X-ray source. Hence a correlation between eclipse depth and $N_H$ will be found, so long as the X-ray source is large enough to be partially covered for the smaller clouds, as seems to be the case (e.g. Risaliti et al. 2009a). X-ray eclipse observations undertaken to map the general relativistic metric via the broad Fe-K line (Risaliti et al. 2011) can be used to measure $f_c$ for the X-ray eclipsers. The X-ray source is not expected to be a uniform disk, though. Special and general relativistic effects distort the X-ray source significantly in the observed plane (e.g. Armitage and Reynolds 2003). Deviations from a simple $N_H$-$f_c$ correlation must then carry information about the X-ray source morphology

For spherical clouds with $n_{10} = 1$ and $10^{22} < N_H < 10^{24.5}$ cm$^{-2}$ (i.e. $\tau_{Compton} = 1$, section 3.4), we find $5\times10^{11} < r_c < 1.5\times10^{14}$ cm. The Schwarzschild radius for a $10^8$ M$_\odot$ black hole is $3\times10^{13}$ cm. Quasar X-ray sources are thought to be at most a few $r_g$ in diameter (Risaliti et al., 2007, Chartas et al. 2009, Emmanoulopoulos et al. 2014). A raindrop cloud with these sizes could eclipse the quasar X-ray source at covering factors from $10^{-4}$ up to total coverage. Higher mass black holes have correspondingly larger X-ray sources (Fender et al. 2007) and should have shallower eclipses with covering factors, $f_c = 10^{-6} - 1$ for $M_8 = 10$. At present probing $f_c < 0.01$ is infeasible.

---

[8] http://global.jaxa.jp/press/2016/04/20160428_hitomi.html

The implied cloud masses are $M_c = 8.75 \times 10^{21} \, n_{10}^{-2} \, N_{22}^{3}$ g, and so span the range $10^{20} < M_c < 3 \times 10^{29}$ g, or $5 \times 10^{-14} < M_c < 1.5 \times 10^{-4} \, M_\odot$. This large range easily encompasses the $\sim 4 \times 10^{-11} \, M_\odot$ estimate of Maiolino et al. (2010) for the head of the cometary clouds and the $4 \times 10^{-10} \, M_\odot$ mass of the tail.

### 4.4 Filling factor and number of BELR clouds

The filling factor of spherical clouds, $\varepsilon$, of the BELR volume is simply $\varepsilon = N_c \, (r_c/R)^3$, where $N_c$ is the number of clouds of radius $r_c$ within $R$ (e.g. Peterson 1997). The value of $\varepsilon$ for the BELR is derived from emission line strengths and reverberation mapping radii. Typical values are $\varepsilon \sim 3 \times 10^{-7}$ (Peterson 1997). The clouds will then be separated by $\gg$ 10 cloud diameters and will act as independently moving clouds.

X-ray eclipses find $r_c \sim 10^{13}$ cm, and $R_{1000} = 1$, $M_8 = 1$, is $3 \times 10^{16}$ cm. For $\varepsilon \sim 3 \times 10^{-7}$ there would then be $\sim 8000$ BELR clouds. However this size includes the greatly elongated "cometary" tail of the clouds. The denser head has a size $r_c \sim 3 \times 10^{11}$ cm (Maiolino et al. 2010). If we assume cylindrical clouds of this radius and of length, $l = 10^{13}$ cm, then $N_c \sim < 10^7$. The cloud volume appropriate to the BELR should be made somewhat smaller as the far part of the tail is seen to have FeXXV-XXVI absorption and so is more highly ionized than the BELR clouds (Maiolino et al. 2010, Risaliti et al., 2010). This is comparable to the Maiolino et al. 2010) estimate of $\leq 10^7$ clouds, but is low compared with the limit of $> 10^8$ clouds from Arav et al. (1998). Bianchi et al. (2012) suggest that the velocity spread in a cometary cloud could greatly reduce the number of clouds needed to match the smoothness of the BEL profiles. In addition, the Arav et al. (1998) estimate includes all H$\alpha$ emitting BELR clouds, including those beyond $R_{1000}$, so allowing more clouds to be accommodated in the model than in the above estimate. Both estimates of BELR cloud numbers remain viable at present.

The implied mass of the BELR is then $> 10^{-6} \, M_\odot$ to $> 10^4 \, M_\odot$ (using the two cloud mass estimates from section 4.3). The latter value is in line with observations (Baldwin et al., 2003).

### 4.5 Quasar Rain and the Baldwin Effect

MCGV showed that enhanced X-rays relative to the UV reduce the effectiveness of line-driving by overionizing the target gas. X-ray-loud quasars would then have smaller values of $\mathcal{M}$, leading to slower accelerations and so more time for clouds to condense. Hence, BELR clouds should be more numerous in such quasars if quasar rain is the source of all BELR clouds, with consequently higher equivalent width emission lines. X-ray loud quasars, especially the low luminosity Seyfert galaxies, should have more clouds and stronger BELs. In principle this effect could lead to the Baldwin effect – that higher luminosity quasars have lower equivalent width CIV BELs (Baldwin 1977, Shields 2007). It is not clear why in this picture the higher ionization potential ions, some of which are thought to be in outflow (Leighly and Moore 2004, but see also Grier et al.

2013), would show the observed stronger Baldwin effect (Dietrich et al. 2002). A quantitative investigation of this possiblitiy would be of interest.

There is a factor ~50 observed range of X-ray to UV flux ratios in quasars, with low optical luminosity quasars being more X-ray loud (Gibson, Brandt & Schneider 2008; Young, Risaliti & Elvis 2010). However, Risaliti and Lusso (2015) and Lusso and Risaliti (2016) show that most of this range is due to the effects of host galaxy contamination and dust reddening in the optical/ultraviolet bands. These effects would have confused any prior attempt to look for BELR trends with $L_X/L_{UV}$.

The cleaned SDSS quasar sample of Lusso and Risaliti (2016), with all reddened, large host galaxy contribution and low signal-to-noise objects removed, spans a range of 1000 in UV luminosity. They find that $L_{UV}$ = const. $L_X^{0.6}$, so $L_X$ only spans a factor ~65, and the range of $L_X/L_{UV}$ in the sample is a factor ~16. This should be sufficient to look for systematic changes in the BELR that may be connected with the quasar rain model.

There is a small subclass of weak line quasars (McDowell et al., 1995, Shemmer and Lieber 2015) which may give clues to the extreme end of the Baldwin effect. So far they do not appear to prefer high luminosities, although they do cluster at high Eddington rates (Shemmer and Lieber 2015).

## 5. CHALLENGES TO THE MODEL

### *5.1 Is there BELR Infall after all?*

The VRRM infall results that required infall in the BELR on the near side of the quasar (Section 1) were the starting point for the quasar rain model.

A challenge to the infall case comes from *XMM-Newton* observations of the narrow line Seyfert 1 Mrk 766. During the three absorption events in this long observation, Risaliti et al. (2010) found high ionization Fe-K XXV and XXVI Kα absorption lines with *outflow* velocities of 3000 - 15,000 km s$^{-1}$. As these lines are not seen outside of eclipsing events the neutral and the highly ionized absorbing gas must be associated. The velocities are inconsistent with their all being the same. Gas at this high ionization level would not produce the optical and UV BELs. However, these lines might arise in the tail of the eclipsing cloud, as in NGC1365. There is insufficient signal-to-noise in the time-resolved spectra of these events to know where the high ionization lines lie in the absorbing clouds. In the quasar rain model the tail is being ablated from the head by a radially flowing wind, which must then accelerate the tail toward the local velocity of the wind. Whether this acceleration is rapid enough to account for the outflow velocities in Mrk 766 without violating the quasar rain model is unclear and needs more detailed modeling.

**X**-ray spectroscopy with a resolving power, $R$~1000 would provide much better signal-to-noise for narrow absorption lines than the $R$~30 of current CCD detectors (Haberl et al. 2002, Garmire et al. 2003), and could resolve infall velocities of 1000 - 2000 km s$^{-1}$ in the Fe-K absorption lines. Line velocity centroids can be measured to a fraction of this value (Hitomi Collaboration 2016). Observing the heads of cometary X-ray eclipsers at that resolving power should show that they are infalling, and should resolve a set of Fe-K absorption lines of increasing ionization along the tails. This will map out the kinematics of the ablated gas and so of the local wind.

The VRRM data is presently mostly based on Hβ. Not all BELs behave the same, however. The CIV BEL is well known to be blueshifted relative to the lower ionization lines such as Hβ (Gaskell 1982, Leighly and Moore 2004). This shift is normally interpreted as outflow in a wind (e.g. Leighly 2004, Richards 2012). How can this outflow be reconciled with BELR inflow? High ionization absorption lines (CIV, OVI) are seen as part of the warm phase of the WA wind (e.g. Mathur et al. 1995, Kriss 2005). It could be that the blueshifted component of CIV is due to this warm phase that is successfully accelerated to $v_{esc}$. A clever alternative to obtain a blueshift from infalling gas was suggested by Gaskell and Goosmann (2013) using scattering off an inflowing medium.

One of the four exemplars of VRRM infall, PG 2130+099 (Grier et al. 2013), has been well observed in both X-rays and the UV. PG 2130+099 has a quite normal X-ray to UV ratio that is unchanged in data taken over 20 years apart (Wilkes & Elvis 1987, Cardaci et al. 2009). A higher X-ray/UV ratio might be expected in an early example, as this would have produced lower wind acceleration and more time for clouds to form (Section 3.1). Also, while the high resolution *XMM-Newton* RGS grating spectrum of PG2130+099 does point to a two-phase X-ray warm absorber in PG2130+099, the low ionization phase has an unusually high temperature of $(1.8\pm0.2) \times 10^5$ K (Cardaci et al. 2009), well above that of the BELR. A lower temperature absorber is seen in CIV in the UV (Crenshaw et al. 1999). Interestingly both Crenshaw et al. (1999) and Cardaci et al. (2009) note the possible presence of inflows. Good UV and X-ray spectra of the other BELR inflow objects, especially of the strongest case, Arp 151 (Pancoast et al. 2014) are needed to test this model.

*5.2 Distant Warm Absorbers?*

Clearly this model can be falsified by observations that place most low ionization WA gas beyond the BELR radius. The location of quasar WAs is presently unclear. While the studies by Krongold and collaborators (see footnote 3), as well as by Netzer et al. (2003), Kriss et al. (2003) and Blustin et al. (2003) tended to give distances close to the BELR, many other studies, mainly using UV lines, favor larger radii (see summary by Crenshaw and Kraemer 2012). The quality of the X-ray data in particular is sufficiently limited, especially in resolving power ($R \sim 400$) leading to major line blending and diluting the predicted variations. The limited signal-to-noise then allows multiple interpretations. In principle metastable X-ray lines can be used as density diagnostics, from which the radius follows directly. But these measurements would require significantly improved X-ray spectroscopy capabilities (e.g. Smith et al. 2016[9], Gaskin et al. 2015[10]).

---

[9] See ARCUS web site: http://www.arcusxray.org/index.html
[10] See **Lynx**/X-ray Surveyor web site: http://cxc.harvard.edu/cdo/xray_surveyor/

# 6. CONCLUSIONS

I have presented a simple model for the origin of the broad emission line region in quasars and AGN that unifies the three manifestations of cool gas: the broad emission line (BEL) clouds, the low ionization phase of a warm absorber accretion disk wind, and X-ray eclipsing clouds. Estimates of the timescales involved are consistent with this mechanism occurring and with the observational constraints. Several tests of the model can be made. The model uses robust physics that will come into play when a supermassive black hole accretes at a reasonable fraction of the Eddington rate.

The scenario is that condensations form naturally and quickly (in days to weeks) at sub-escape velocities in the warm accretion disk wind found in most quasars. The rapid variability of the X-ray continua in quasars will initiate cloud collapse and will perturb the WA gas into stable regions of the thermal equilibrium curve, allowing the cool phase to form. The thermal instability that produces the raindrops is especially pronounced if the gas has super-Solar abundances, as is normal in quasars and AGN, but always spans only a factor 2-3 in pressure and temperature. Hence constant density BELR clouds are expected and can explain the observed $L_{bol}^{1/2}$ scaling of the BELR radius. These ~$10^4$ K condensations will have the physical conditions seen in BEL clouds.

The condensed clouds have $N_H > 10^{22}$ cm$^{-2}$ up to at least Compton thick and perhaps $10^{25} - 10^{26}$ cm$^{-2}$. They have ~100 times higher densities compared to the WA wind. This drops their force multiplier to unity, so they can no longer be radiatively accelerated. There will be $>10^8$ of these clouds, with a total mass that could be several 1000 M$_\odot$. These clouds will stall in the wind and will rain back down toward the accretion disk on a dynamical timescale of order a year. These infalling BEL clouds may then produce the infall signatures seen in velocity resolved reverberation mapping.

These "raindrops" rapidly become highly supersonic (Mach ~10 – 100) relative to the warm absorber outflow and will be destroyed during their first orbit on a months timescale, which is consistent with both VRRM and X-ray eclipse observations. Narrow (~0.6 - 6 deg) Mach cones will form behind the clouds, consistent with the shapes seen in the X-ray eclipses of NGC 1365. Those quasars with observed BEL inflows should be searched for UV or X-ray absorption by low ionization gas, especially for X-ray eclipses.

This scenario of "the wind and the rain" has some appeal. Theoretically it retains the simplicity of a quasar accretion disk wind explanation for virtually all the emission and absorption line features in quasars without introducing another component. Observationally it naturally explains the details of the "cometary tails" seen in X-ray eclipses, the elliptical orbits common in the BELR, the (luminosity)$^{1/2}$ scaling of BELR radii and, possibly, the Baldwin effect.

Only a first sketch of the quasar rain model has been presented here, based on simple timescale arguments. Detailed modeling is needed to understand the range of parameter space in which quasar rain can occur, and to test the model. These investigations include:

1. How $\tau_{cool}$ is changed in a photoionized gas (section 3.1);
2. Whether X-ray/UV variability can drive WA gas to stable branches on the S-curves (section 3.2);
3. How the S-curves are altered when $\tau_X \sim \tau_{cool}$ (section 3.2);

4. Whether large clouds can reach $v_{esc}$ and so escape to form the narrow emission line region (section 3.4); and
5. Whether the Baldwin effect is explained by the larger $\tau_{esc}/\tau_{cool}$ ratio in X-ray louder, high luminosity quasars.

If the basic argument holds, then a full hydrodynamical treatment would be worthwhile. One output would be an estimate of the BELR flattening (section 2.3).

As noted by Tepper-Garciá et al. (2015) in the context of the Galactic halo, rain that evaporates before reaching the ground is called *virga*. As the infalling quasar clouds are destroyed before reaching the accretion disk, *quasar virga* may be a more precise name for quasar rain.

I thank Anna Pancoast and Tommaso Treu for stimulating discussions that led to this work, Susmita Chakravorty for sharpening the argument about multi-phase media, Paul Nulsen and Dan Patnaude for helping with cloud destruction timescales, Guido Risaliti and Jonathan McDowell for clarifying the cloud dynamics, Frank van den Bosch for a note on equilibrium timescales, and Norm Murray for the escape time calculation. Mike Goad pointed out an optical depth issue for the raindrops. The referee, Emanuele Nardini, provided excellent comments multiple times that helped me escape embarrassment and led to a many key improvements in the paper. Any remaining errors are, clearly, my own. I thank the Aspen Center for Physics, operated under NSF grant # 1066293, for hosting me while this paper was developed. Lastly, I recall Mike Penston who always used to ask "What's the weather like in NGC4151 today?" With both wind and rain in quasars, Mike may have been more prescient than he knew.

## REFERENCES

Andrade-Velázquez M., et al., 2010, ApJ, 711, 888.

Alexander, T., and Netzer, H., 1994, MNRAS, 270, 781.

Antonucci, R.R.J. and Miller, J.S., 1985, ApJ, 297, 621.

Arav, N., Barlow, T.A., Laor, A., Sargent, W.L.W., & Blandford, R.D., 1998, MNRAS, 297, 990.

Arav, N., Borguet, B., Chamberlain, C., Edmonds, D., and Danforth, C., 2013, MNRAS, 436, 3286.

Armitage, P.J. and Reynolds C.S., 2003, MNRAS, 341, 1041.

Baldwin, J.A., 1977, ApJ, 214, 679.

Baldwin, J.A., Ferland, G.J., Korista, K.T., Hamann, F., and Dietrich, M., 2003, ApJ, 582, 590.

Baskin, A., Laor, A. and Stern, J., 2014a, MNRAS, 438, 604.

Baskin, A., Laor, A. and Stern, J., 2014b, MNRAS, 445, 3025.

Bentz, M.C., Peterson, B.M., Pogge, R.W., Vestergaard, M. & Onken, C.A., 2006, ApJ, 644, 133.

Bentz, M.C., Horne, K., Barth, A. J., et al., 2010, ApJL, 720, 46.

Bentz, M.C., et al., 2013, ApJ, 767, 149.

Bianchi, S., Maiolino, R., and Risaliti, G., 2012, Advances in Astronomy, 2012, id.78230.
Blumenthal, G.R., and Mathews, W.G., 1979, ApJ, 233, 479.
Blustin, A.J. et al., 2003, ApJ, 598, 232.
Blustin, A.J. et al., 2004, Advances in Space Research, 34, 2561.
Bowler, R.A.A., et al., 2014, MNRAS, 445, 359.
Braito, V., Reeves, J.N., Gofford, J., Nardini, E., Porquet, D., and Risaliti, G., 2014, ApJ, 795, 87.
Cardaci, M.V., et al. 2008, A&A, 505, 541.
Castor, J.I., Abbott, D.C. and Klein, R.I., 1975, ApJ, 195, 157.
Chakravorty, S., et al. 2008, MNRAS, 384, 24.
Chakravorty, S., Kembhavi, A.K, Elvis, M., and Ferland, G., 2009, MNRAS, 393, 83.
Chakravorty, S., et al. 2012, MNRAS, 422, 637.
Chartas, G., Kochanek, C.S., Dai, X., Poindexter, S., and Garmire, G., 2009, ApJ, 693, 174.
Crenshaw, D.M. and Kraemer, S.B., 1999, ApJ, 521, 572.
Crenshaw, D.M. and Kraemer, S.B., 2012, ApJ, 753, 75.
Dietrich, M., et al., 2002, ApJ, 581, 912.
Dehnen, W. and King, A.R., 2013, ApJ, 777, L28.
Elvis M., et al. 1994, ApJS, 95, 1.
Elvis M., 2000, ApJ, 545, 63.
Elvis M., Risaliti, G., Nicastro, F., Miller, J.M., Fiore, F., & Puccetti, S., 2004, ApJ, 615, L25.
Emmanoulopoulos, D., Papadakis, I.E., Dovčiak, M., and McHardy, I.M., 2014, MNRAS, 439, 3931.
Emmering, R.T., Blandford, R.D., and Shlosman, I., 1997, ApJ, 385, 460.
Fender, R., et al., 2007, arXiv:0706.3838.
Ferland, G.J., et al., 1992, ApJ, 387, 95.
Fields, D. L., Mathur, S., Krongold, Y., Williams, R., & Nicastro, F. 2007, ApJ, 666, 828.
Fischer, T.C., Crenshaw, D.M., Kraemer, S.B., Schmitt, H.R., Mushostzky R.F., and Dunn, J.P., 2011, ApJ, 727, 71.
Fischer, T.C., Crenshaw, D.M., Kraemer, S.B., and Schmitt, H.R., 2011, ApJ, 209, 1.
Frank, J., King, A. & Raine, D.J., 2002, "Accretion Power in Astrophysics", 3$^{rd}$ edition, [Cambridge UK: Cambridge University Press].
Fromerth, M.J., and Melia, F., 2001, ApJ, 549, 205.
Garmire, G., et al., 2003, SPIE, 4851, 28.
Gaskell, C.M., 1982, ApJ, 263, 79.
Gaskell, C.M., 2009, NewAR, 53, 140.
Gaskell, C.M. and Goosmann, R.W., 2013, ApJ, 769, 30.
Gaskin, J.A., et al., 2015, 2015SPIE.9601E..0JG
Gaspari, M., Brighenti, F. and Temi, P., 2015, A&A, 579, A62.
Gehrels, N., & Williams, E.D., 1993, ApJ, 418, L25.

Gibson, R.R., Brandt, W. N., Schneider, D.P., 2008, ApJ, 685, 773.

Gibson, R.R., et al., 2009, ApJ, 692, 758.

Giommi, P., Padovani, P., Polenta, G., Turriziani, S., D'Elia, V., and PIranomonte, S., 2012, MNRAS, 420, 2899.

Grier, C.J., et al. 2013, ApJ, 764, 47.

Groves, B.A., Heckman, T.M., and Kauffmann, G., 2006, MNRAS, 371, 1559.

Haberl, F., Briel, U.G., Dennerl, K. and Zavlin, V.E., 2002, arXiv:0203235.

Hamann, F. and Ferland, G.J., 1999, Ann.Rev. A&A, 37, 487.

Hamann, F., Warner, C., Dietrich, M., and Ferland, G., 2007, ASP Conference series, 373, 653.

Hao, H., et al., 2013, MNRAS, 434, 3104.

Heitsch, F., and Putman, M.E., 2009, ApJ, 698, 1485.

Hitomi Collaboration, 2016, Nature, 535, 117.

Holczer, T., Behar, E., and Kaspi, S., 2007, ApJ, 663, 799.

Hopkins, P.F. and Elvis, M., 2010, MNRAS, 401, 7.

Isler, J.C., et al., 2015, ApJ, 804, 7.

Johnson, C.I., Rich, M.R., Kobayashi, C., Kunder, A., and Koch, A., 2014, AJ, 148, 76.

Klein, R.I., McKee C.F., and Colella, P., 1994, ApJ, 420, 213.

Kollmeier, J., et al., 2006, ApJ, 648, 128.

Korista, K., Baldwin, J., Ferland, G. and Verner, D., 1997, ApJS, 108, 401.

Kormendy, J. and Ho, L.C., 2013, ARA&A, 51, 511.

Kriss, G.A., 2003, A&A, 403, 473.

Kriss, G.A., 2005, ASP Conf. Series, 348, 499.

Krolik, J.J., 1977, Physics of Fluids, 20, 364.

Krolik, J.J., 1988, 325, 148.

Krolik, J.H., McKee, C.F. and Tarter, C.B., 1981, ApJ, 249, 422.

Krongold, Y., Nicastro, F., Brickhouse, N. S., Elvis, M., Liedahl, D. A., & Mathur, S. 2003, ApJ, 597, 832.

Krongold, Y., Nicastro, F., Brickhouse, N. S., Elvis, M., & Mathur, S. 2005a, ApJ, 622, 842.

Krongold, Y., Nicastro, F., Elvis, M., Brickhouse, N. S., Mathur, S., & Zezas, A. 2005b, ApJ, 620, 165.

Krongold, Y., Nicastro, F., Elvis, M., Brickhouse, N., Binette, L., Mathur, S., & Jiménez-Bailón, E. 2007, ApJ, 659, 1022

Krongold, Y., et al. 2009, ApJ, 690, 773.

Krongold, Y., et al. 2010, ApJ, 710, 360.

Laha, S., Guainazzi, M., Dewangan, G.C., Chakravorty, S., & Kembhavi, A.K., 2014, MNRAS, 441, 2613.

Lamers, H.J.G.L.M., and Cassinelli J.P., 1999, Introduction to Stellar Winds [CUP:Cambridge].

Laor, A., Barth, A.J., Ho, L.C., and Filippenko, A.V., 2006, ApJ, 636, 83.

Leighly, K., 2004, ApJ, 611, L125.
Leighly, K.M., and Moore, J.R., 2004, ApJ, 611, 107.
Leighly, K., & Moore, J.R., 2004, ApJ, 611, L107.
Lucy, L.B. & White, R.L., 1980, ApJ, 241, 300.
Lusso, E., and Risaliti, G., arXiv:1602.01090.
MacGregor, K.B., Hartmann, L., and Raymond, J.C., 1979, ApJ, 231, 514.
Maiolino, R., et al. 2010, A&A, 517, A47.
Marconi, A., et al., 2008, ApJ., 678, 693.
Mathews, W.G. & Ferland, G.J., 1987, ApJ, 323, 456.
Mathews, W.G. and Doane, J.S., 1990, ApJ, 352, 423.
Mathur, S., Elvis, M., and Wilkes, B., 1995, ApJ, 452, 230.
McDowell, J.C., et al., 1995, ApJ, 450, 585.
M$^c$Hardy, I.M., et al., 2006, Nature, 444, 730.
M$^c$Hardy, I.M., et al., 2014, MNRAS, 444, 1469.
Misuda, K., et al., 2014, Proceedings of the SPIE, Volume 9144, id. 91442A.
Mortlock, D.J., et al., 2011, Nature, 474, 616.
Murray, N., Chiang, J., Grossman, S.A., and Voit, G. M. 1995, ApJ, 451, 498. (MCGV)
Murray, N., and Chiang, J. 1995, ApJ, 454, L105.
Murray, N., and Chiang, J. 1998, ApJ, 494, 125.
Nagao, T., et al., 2004, ApJ, 128, 109.
Nagao, T., Maiolino, R., and Marconi, A., 2006, A&A, 447, 863.
Nicastro, F., Fiore, F., Perola G.C., & Elvis M., 1999, ApJ, 512, 184.
Netzer, H., 1975, MNRAS, 171, 395.
Netzer, H., et al. 2003, ApJ, 599, 933.
Netzer, H., 2013 "The Physics and Evolution of Active Galactic Nuclei" [Cambridge: Cambridge University Press]
Ohsuga, K., and Mineshige, S., 2014, Space Science Reviews, 183, 353.
Osterbrock, D.E., and Ferland, G.J., 2006, "Astrophysics of Gaseous Nebulae and Active Galactic Nuclei", 2nd. ed. [Sausalito, CA: University Science Books].
Pancoast, A., et al., 2014, MNRAS, 445, 3073.
Pâris, I., et al., 2014, A&A, 563, 54.
Patnaude, D.J., and Fesen, R.A., 2014, ApJ, 789, 138.
Perry, J.J. and Dyson, J.E., 1985, MNRAS, 213, 665.
Peterson, B.M., 1997, An Introduction to Active Galactic Nuclei [Cambridge:Cambridge] ISBN 0 521 47922 8.
Peterson, B.M., and Wandel, A., 1999, ApJ, 521, L95.
Peterson, B.M., et al., 2005, ApJ, 613,682.
Pian, E., et al., 1999, ApJ, 521, 112.
Piconcelli, E., Jimenez-Bailón, E., Guainazzi, M., Schartel, N., Rodríguez-Pascual, P. M., & Santos-Lleó, M., 2005, A&A, 432, 15.
Pittard, J.M., Dyson, J.E., Falle, S.A.E.G. and Hartquist, T.W., 2001, A&A, 375, 827.

Popović, L, Č., 2003, ApJ, 599, 140.
Puccetti, S., Fiore, F., Risaliti, G., et al. 2007, MNRAS, 377, 607.
Proga, D. 2003, ApJ, 585, 406.
Rees, M.J., 1987, MNRAS, 228, 47.
Risaliti, G., Elvis, M., Fabbiano, G., Baldi, A., & Zezas, A. 2005, ApJ, 623, L93.
Risaliti, G., Elvis, M., Fabbiano, G., Baldi, A., Zezas, A., and Salvati, M., 2007, ApJ, 659, L111.
Risaliti, G., Salvati, M., Elvis, M., et al. 2009a, MNRAS, 393, L1.
Risaliti, G., Miniutti, G., Elvis, M., et al. 2009b, ApJ, 696, 160.
Risaliti, G. and Elvis M., 2010, A&A, 516, A89.
Risaliti, G., et al., 2011, MNRAS, 417, 178.
Risaliti, G., and Lusso, E., 2015, ApJ, 815, 33.
Schmidt, M., 1963, Nature, 197, 1040.
Scoville, N., and Norman, C., 1988, ApJ, 332, 163.
Shields, G.A., 1977, Astrophysical Letters, 18, 119.
Shields, J., 2007, ASP Conference Series, 373, 355.
Steinhardt, C.L., & Elvis M., 2010, MNRAS, 402, 2637.
Shakura, N.I., and Sunyaev, R.A., 1973, A&A, 24, 337.
Shemmer, O., and Lieber, S., 2015, ApJ, 805, 124.
Smith, R.K., et al., 2016, 2016AAS...22714728S.
Steenbrugge, K.C., Kaastra, J.S., de Vries, C.P., & Edelson, R. 2003, A&A, 402, 477.
Steenbrugge, K.C., et al., 2005, A&A, 434, 569.
Tepper-Garciá, T., Bland-Hawthorn, J., and Sutherland, R.S., 2015, ApJ, 813, 94.
Torricelli-Ciamponi ,G., Pietrini, P., Risaliti G., & Salvati, M., 2014, MNRAS, 42, 2116.
Tucker, W.H., 1975, Radiation Processes in Astrophysics [MIT:Cambridge MA]
Voit, G.M. & Donahue, M., 2015, ApJ Letters, 799, L1.
Young, M., Elvis, M., & Risaliti, G., 2010, ApJ, 708, 1388.
Vanden Berk, D.E., et al., 2004, ApJ, 601, 692.
Vilkoviskij, E.Y., and Czerny, B., 2002, A&A, 387. 804.
Vietri, M., Ferrara, A., and Miniati, F., 1997, ApJ, 483, 262.
Waters, T., and Proga, D., 2016, arXiv:1603.01915.
Waters, T., et al., 2016, arXiv:1601.05181.
Wilkes, B.J. and Elvis, M., 1987, MNRAS, 323, 243.
Wills, B.J., & Browne, I.W.A., 1986, ApJ, 302, 56.